\documentclass[12pt]{iopart}

\usepackage{graphicx} 
\begin{document}

\title[A.Szczurek et al.]{
 The effect of the spectator charge
on the charged pion spectra in peripheral
ultrarelativistic heavy-ion collisions.
}

\author{A Szczurek$^{1,2}$, A Rybicki$^1$, A Z G\'orski$^1$}

\address{$^1$ Institute of Nuclear Physics 
ul. Radzikowskiego 152, 31-342 Krak\'ow, Poland}
\address{$^2$ University of Rzesz\'ow, ul. Rejtana 16, 
35-959 Rzesz\'ow, Poland}
\ead{antoni.szczurek@ifj.edu.pl}
\begin{abstract}
We estimate the electromagnetic effect of the spectator charge on the
momentum spectra of charged pions produced in peripheral Pb+Pb
collisions at SPS energies. We find a large effect which results
in strongly varying structures in the $x_F$ dependence of the
$\pi^+/\pi^-$ ratio, especially at low transverse momenta where a 
deep valley in the above ratio is predicted at $x_F \sim$ 
0.15 -- 0.20.
The effect depends on initial conditions. Thus, it
provides new information on the space and time evolution of the 
non-perturbative pion creation process.
\end{abstract}


\section{Introduction}

The nuclear collisions at SPS energies, when combined with the NA49
apparatus, open a possibility to study several subtle effects \cite{mul}
which are difficult to study at larger energies, like e.g. at RHIC.
On the other hand, a few new typically nuclear effects have been
predicted recently for SPS energies \cite{SB04,PS04}.

Recently the NA49 experiment made a preliminary observation \cite{isosp} of
a new interesting phenomenon in peripheral collisions, which after a
more refined experimental analysis was advocated
\cite{meson06} to be a Coulomb effect due to spectator charge.
Here we discuss the origin of this effect.
In particular, we discuss electromagnetic (EM) interaction between the
remnants of the two colliding nuclei and $\pi^+$ and $\pi^-$
produced in the collision.  
The two highly charged spectator systems 
generate a rapidly changing EM field which modifies the
pion trajectories. This causes a distorsion of 
observed kinematical pion spectra.  
We show that this distorsion is interrelated to the dynamics of
the collision, and in particular to the time evolution and initial
conditions of the participant and spectator zones.

We study this EM effect for the specific case of peripheral Pb+Pb
collisions at SPS energies (158 GeV/nucleon beam energy,
$\sqrt{s}_{NN}$=17 GeV). 
This short presentation is based on our recent detailed studies \cite{RS06}.

\section{Propagation of pions in the EM field}

Our approach is based on a Monte Carlo simulation of initial conditions
and subsequent propagation of charged pions in the EM
field of moving spectator charge. In Ref.\cite{RS06} we have devoted
a separate section where we discuss the choice of the initial conditions.
These roughly correspond to experimental samples available at the SPS.
Here we concentrate only on propagation of pions in the EM field.

We define $\vec{E}'_{L}$ as the constant electrostatic field generated
by the spectator $L$ in its rest frame, and $\vec{E}''_{R}$ as the field 
generated by the spectator $R$ in its rest frame. 
We assume both spectator systems to be uniform spheres with a normal
nuclear density $\rho=0.17$/fm$^{3}$ and with a total charge $Q=70$
elementary units. Then the static electric fields is given
by a simple nuclear-physics text-book formula \cite{RS06}.

We transform the fields $\vec{E}'_{L}$,
$\vec{E}''_{R}$ to the overall center of mass system.
Here
\begin{equation}
\vec{E}_{L}(\vec{r},t) = \gamma_s \vec{E}'_{L}
(\vec{r'_c})
 - \frac{\gamma_s^2}{\gamma_s+1}
\; \frac{\vec{v}_{L}}{c} \; \left( \frac{\vec{v}_{L}}{c} \cdot 
\vec{E}'_{L}
(\vec{r'_c})
\right)
,
\vec{B}_{L}(\vec{r},t) = \gamma_s \left( \frac{\vec{v}_{L}}{c} \times 
\vec{E}'_{L} 
(\vec{r'_c})
\right)
\end{equation}
for the left spectator and
\begin{equation}
\vec{E}_{R}(\vec{r},t) = \gamma_s \vec{E}''_{R}
(\vec{r''_c})
 - \frac{\gamma_s^2}{\gamma_s+1}
\; \frac{\vec{v}_{R}}{c} \; \left( \frac{\vec{v}_{R}}{c} \cdot 
\vec{E}''_{R}
(\vec{r''_c})
\right)
,
\vec{B}_{R}(\vec{r},t) = \gamma_s \left( \frac{\vec{v}_{R}}{c} \times 
\vec{E}''_{R} 
(\vec{r''_c})
\right)
\end{equation}
for the right spectator. 

In the equations above, the $\gamma_s$ factor is defined as 
$\gamma_s=(1-v^2_s/c^2)^{-1/2}$. The vectors $\vec{E}_{L}$ $(\vec{E}_{R})$ 
and $\vec{B}_{L}$ $(\vec{B}_{R})$ are respectively the electric and 
magnetic fields generated by the left (right) spectator at the space-time 
position $(\vec{r},t)$.

The resulting pion trajectory $\vec{r}_{\pi}(t)$ is 
defined by its time-dependent velocity $\vec{v}_\pi(\vec{r},t)$:
\begin{equation}
\frac{d \vec{r}_{\pi}}{d t} = \vec{v}_{\pi}(\vec{r},t) = 
\frac{\vec{p}_\pi \; c^2}{\sqrt{p^2_\pi+m^2_\pi }} \; ,
\label{velocity}
\end{equation}
where $m_\pi$ is the pion mass.

\section{Results}

In this Section we present some results of our Monte-Carlo studies
described in detail in \cite{RS06}.

\begin{figure}               
\begin{center}
\includegraphics[height=4cm]{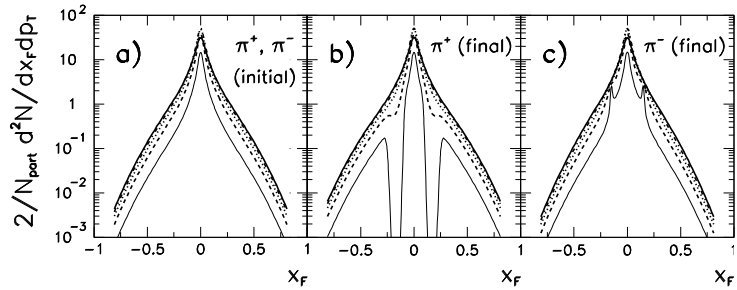}
 \caption{Double-differential density of $\pi^+$ and $\pi^-$
produced per participant pair in peripheral
Pb+Pb reactions.
 {\bf a)}~Initial density of emitted $\pi^+$ and $\pi^-$.
 {\bf b)}~Density of $\pi^+$ in the final state.
 {\bf c)}~Density of $\pi^-$ in the final state. 
 The pion density is drawn as a function of $x_F$ at $p_T=25$
MeV/c (thin solid), 75 MeV/c (dash), 125 MeV/c (dot), 175 MeV/c
(dash-dot), and 325 MeV/c (thick solid); $p_T$ values
corresponds to a bin of $\pm 25$~MeV/c. This simulation was made
for $t_E$=0.
 \label{dndx}
}
\end{center}
\end{figure}

Our main results are illustrated in Fig.~\ref{dndx}. Panel {\bf (a)} shows
the initial spectra of emitted pions. As explained in Ref.\cite{RS06},
in our simple model these spectra are identical for $\pi^+$ and $\pi^-$.  
The presented $\frac{d^2N}{dx_Fdp_T}$ density distributions are scaled down
by the number of participant pairs. 

In panel {\bf (b)}, the corresponding distributions of $\pi^+$ in the
{final state} of the Pb+Pb reaction are shown. These are obtained by our
Monte-Carlo simulation. It is clearly
apparent that the distributions are distorted by the Coulomb repulsion
between the pion and spectator charges. The effect is largest for pions
moving close to spectator velocities ($x_F\approx\pm 0.15$) and at low
transverse momenta ($p_T=25$~MeV/c). Here, two deep valleys in the $\pi^+$
density are visible. A similar but smaller distorsion is also apparent at
$p_T=75$~MeV/c.

An opposite distorsion is present for $\pi^-$ densities shown in panel 
{\bf (c)}. Negative pions are attracted by the positive spectator charge 
and gather at low transverse momenta close to spectator 
velocities. This results in the presence of two large peaks at 
$x_F\approx\pm 0.15$. Remnants of these peaks are apparent at 
$p_T=75$~MeV/c.

\begin{figure}             
\begin{center}
\includegraphics[height=4cm]{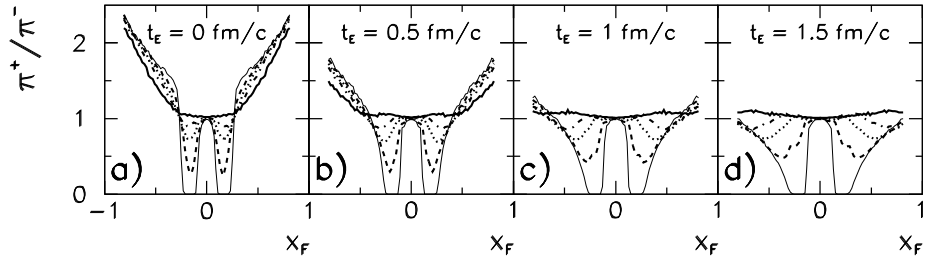}
\caption{Ratio of cross sections for $\pi^+$ and $\pi^-$ 
for a peripheral Pb+Pb reaction, for five values of $t_E$.
The $\pi^+/\pi^-$ 
ratio is drawn as a function of $x_F$ at $p_T=25$ MeV/c (thin solid), 75 
MeV/c (dash), 125 MeV/c (dot), 175 MeV/c (dash-dot), and 325 
MeV/c (thick solid).
 \label{rat}}
 \end{center}
 \end{figure}

Fig.~\ref{rat} shows the $x_F$-dependence of the
${\pi^+/\pi^-}$ for different pion emission time $t_E$
(the time is measured since the time of the closest approach of two colliding
nuclei). For $t_E=$ 0
the spectator Coulomb field appears to produce a characteristic,
complex pattern of deviations from unity. The first element of this
pattern is a double, two-dimensional valley which covers the low-$p_T$
region in the vicinity of $x_F\approx\pm 0.15$; the valley remains still
visible at $p_T=175$~MeV/c. The second element is a smooth rise of the 
${\pi^+/\pi^-}$ ratio at higher $|x_F|$. This rise is present for 
all the considered $p_T$ values; at fixed $x_F$, the ratio slowly 
decreases with increasing $p_T$.

The central result of our analysis is the sensitivity of spectator-induced 
Coulomb effects to initial conditions.
Indeed, clear differences appear for different $t_E$ values
(Figs.2a,b,c,d). With 
increasing $t_E$, a broadening of the double valley at $x_F\approx\pm 
0.15$, and a decrease of the ${\pi^+/\pi^-}$ ratio at higher absolute 
$x_F$, are visible.

\section{Conclusions}

The electromagnetic interaction between highly charged spectators
and charged pions produced in the peripheral Pb+Pb reaction has been
studied by means of a simplified but realistic model. This interaction
leads to significant distorsions on the final state densities of
$\pi^+$ and $\pi^-$.

The main feature of this ``Coulomb'' effect is a big dip in the $\pi^+$ 
density distribution at low transverse momenta in the vicinity of 
$x_F\approx\pm 0.15$, accompanied by an increase of $\pi^-$ density
in the corresponding region of phase space.
This results in the presence of a two-dimensional valley in $(x_F,p_T)$
for the $\pi^+/\pi^-$ cross section ratio. In addition, at higher
absolute $x_F$, a smooth increase of $\pi^+/\pi^-$ with $x_F$ may appear.

The sensitivity of this EM effect to initial conditions 
has been estimated. The effect is clearly sensitive to initial
conditions. Changes of the pion emission time by 0.5 
fm/c (in c.m.s. time) are sufficient to modify the observed distorsion 
pattern. In our model, such changes are equivalent to changes of position 
of the formation zone by 0.5 fm relative to the two spectator systems. 
Thus, the EM effect appears to depend on the evolution of 
the pion production process in space and in time. Therefore, it
constitutes a new source of information on the non-perturbative
process of light--meson production.

\subsection{Acknowledgments.}
This work was supported by the Polish State Committee for Scientific
Research under grant no. 1 P03B 097 29.

\end{document}